\documentclass{pasj00}
\draft

\newcommand{\cjaa}{Chinese J. Astron. Astrophys, }
\newcommand{\actaa}{AcA, }

\begin{document}
\SetRunningHead{Author(s) in page-head}{Running Head}

\title{Physical properties and evolutionary status of the W-subtype contact binary V502 Oph with a stellar companion}



%
 \author{Zhou \textsc{Xiao}\altaffilmark{1,2,3}
         Qian  \textsc{Shengbang}\altaffilmark{1,2,3}
         Huang \textsc{Binghe}\altaffilmark{1,2,3}
         Li \textsc{Hao}\altaffilmark{1,3}
         Zhang \textsc{Jia}\altaffilmark{1,2}}
\altaffiltext{1}{Yunnan Observatories, Chinese Academy of Sciences, PO Box 110, 650216 Kunming, China}
 \email{zhouxiaophy@ynao.ac.cn}
\altaffiltext{2}{Key Laboratory of the Structure and Evolution of Celestial Objects, Chinese Academy of Sciences, PO Box 110, 650216 Kunming, China}
\altaffiltext{3}{Graduate University of the Chinese Academy of Sciences, Yuquan Road 19, Sijingshang Block, 100049 Beijing, China}


\maketitle

\begin{abstract}
Multi-color ($B$ $V$ $R_c$ $I_c$) CCD photometric light curves of the contact binary V502 Oph are analyzed using the Wilson-Devinney (W-D) program. The solutions reveal that V502 Oph is a W-subtype contact ($f = 35.3\,\%$) binary system. The temperature difference between its two components is $240K$ and the more massive star has a lower surface temperature. A cool spot is added in our model to account for the light curves' asymmetry (O'Connell effect) and third light is detected for the first time in the light curves' modeling. Combining the orbital inclination ($i = 76.4^{\circ}$) with the published mass function of V502 Oph, the absolute physical parameters of the two components are determined, which are $M_1= 0.46(\pm0.02)M_\odot, M_2=1.37(\pm0.02)M_\odot, R_1=0.94(\pm0.01)R_\odot, R_2=1.51(\pm0.01)R_\odot, L_1=1.13(\pm0.02)L_\odot$ and $L_2=2.49(\pm0.03)L_\odot$. The formation and evolutionary status of V502 Oph are discussed. All photoelectric and CCD times of light minimum about V502 Oph are gathered and its orbital period variations are analyzed. The results show that the orbital period of V502 Oph is decreasing continuously at a rate of $dP/dt=-1.69\times{10^{-7}}day\cdot year^{-1}$, which corresponds to a conservative mass transfer rate of $\frac{dM_{2}}{dt}=-3.01\times{10^{-8}}M_\odot/year$. The light-travel time effect (LTTE) is due to the presence of a close-in tertiary component with a period of $P_3=18.7$ years and an amplitute of 0.00402days. V502 Oph is an ideal target to test the formation and evolution theories of binary and multiple systems in which the light curves, $O - C$ curve and spectroscopic observations are comprehensively researched.
\end{abstract}

\section{Introduction}

V502 Oph (HD 150484, HIP 81703, BD+00 3562, $V$ = $8^{m}.59$) is a W UMa type contact binary system which was discovered to be a variable star by \citet{1935AN....255..401H}. In the following decades, many remarkable work have been done on this target. Spectroscopic observations and radial velocity studies have been carried out by \citet{1948ApJ...108..497S}, \citet{1959ApJ...130..789S}, \citet{1984MNRAS.209..645K}, \citet{2004AJ....127.1712P}; \citet{2007ChJAA...7..558K} and \citet{2011CaJPh..89.1035K}. The analysis revealed that the mass ratio was $q = 0.335(9)$, and the spectral type were G2 V for the more massive component and F9V for the less massive one. Photometric observations and solutions of V502 Oph have been published by  \citet{1967AJ.....72.1028W}; \citet{1969AJ.....74..218B}; \citet{1973IBVS..842....1V}; \citet{1975PZP.....2..161P}; \citet{1982AAS...49..123M}; \citet{1988IBVS.3218....1Z}; \citet{1988AAS...74..265R}; \citet{1998vsr..conf..184R} and \citet{2011MNRAS.412.1787D}. The light curves change over time and may indicate variable surface activity, which made it an ideal target to study the physical mechanism of O'Connell effect \citep{2015IBVS.6127....1K}. The orbital period of V502 Oph was found to change continuously, according to the investigation by \citet{1958BAN....14..131K}; \citet{1971AcA....21..365K}; \citet{1976PASP...88..531S}; \citet{1979IBVS.1569....1M}; \citet{1989RMxAA..17...97H}; \citet{1992AJ....103.1658D}; \citet{2006Ap&SS.304...67Y} and \citet{2006ChJAA...6..331L}.

Most W UMa stars consist of solar-type components with orbital periods of $0.25 d < P < 0.7 d$ and share a common envelope between its two components. They are recognized by continuous brightness variations and nearly equal depth for primary and secondary eclipse. W UMa type binaries are superb laboratories for the study of the formation and evolution scenario of close binaries. The research will provide valuable information on the late stage of stellar evolution connected with the processes of mass transfer, angular momentum loss(AML) and merging of stars, etc \citep{2014AJ....148...96L,2015AJ....150...69Y}. In addition, surveys of companions around eclipsing binaries have generally shown that tertiary components are quite common in W UMa systems \citep{1973A&AS...12....1F,1990BAICz..41..231M,1992PASP..104..663C,1996A&AS..120...63B,2006AJ....131.2986P,2006AJ....132..650D}. The close-in tertiary component may play an important role in the formation and evolution of close binary by shortening the orbital period of the binary system through drawing angular momentum from the central binary system \citep{2006A&A...450..681T}. And also, many different types of objects have been reported to be companions of binary stars, such as stellar companion \citep{2015AJ....150...83Z}, circumbinary planet \citep{2012MNRAS.422L..24Q}, brown dwarf \citep{2009ApJ...695L.163Q} and black hole \citep{2008ApJ...687..466Q}.

As for V502 Oph, lines of a third component were detected while spectroscopic observations were carried out \citep{1998ApJ...504..978H,2006AJ....131.2986P}. The light-travel time effect (LTTE) appearing on the observed-calculated ($O-C$) curve also confirmed the existence of a close-in tertiary component around the binary system \citep{2006Ap&SS.304...67Y,2006ChJAA...6..331L}. Unfortunately, the former photometric solutions had not given any information about third light ($l_3$). Thus, new multi-color high accuracy CCD light curves are still need. In the present paper, $B$ $V$ $R_c$ and $I_c$ bands light curves of V502 Oph are obtained and analyzed using the Wilson-Devinney (W-D) program (Version 2013). More times of light minimum which cover a much longer time scale are collected and fitted. We combine photometric solutions, $O - C$ curve analysis and spectroscopic results together to give out our new research results on V502 Oph, which will shed new insight to our understanding on the formation and evolution theories of binary and multiple systems.

\section{New Multi-color Photometric Observations of V502 Oph}

The $B$ $V$ $R_c$ $I_c$ bands light curves (LCs) of V502 Oph were observed over 11 nights between June 29 and July 18, 2010, using a 20-cm Meade LX200GPS telescope at the Truman State University Observatories in Kirksville, Missouri. An SBIG ST-9XE CCD camera with Johnson $BVR_cI_c$ filters was attached to the telescope \citep{2015IBVS.6127....1K}. Astronomers Control Panel (ACP) was used to communicate between the telescope, CCD camera, focuser, and observatory dome. The telescope and CCD camera were controlled by Maxim DL, which was also used to analyze the observed images and obtain the light curves. A few lines of the original light curve data are displayed in Table \ref{data}.
The resulting differential light curves for V502 Oph in $B$, $V$, $R_c$, and $I_c$ filters are shown in Fig. 1.
The observational time (HJD) were converted to phases with the following linear ephemeris:
\begin{equation}
Min.I(HJD)=2456092.9876+0^{d}.45339\times{E}.\label{linear ephemeris}
\end{equation}

\begin{table}[!h]
\scriptsize
\caption{The original data of V502 Oph}\label{data}
\begin{center}
\begin{tabular}{ccccccccccccc}\hline
  JD(Hel.)       &     Phase   &   $\Delta$m  &    JD(Hel.)       &     Phase   &   $\Delta$m  &     JD(Hel.)       &     Phase    &  $\Delta$m   \\\hline
  2455376.6160   &   0.966034  &   -1.153000  &    2455376.6400   &   0.018968  &   -1.092000  &     2455376.6720   &   0.089548   &  -1.327000   \\
  2455376.6190   &   0.972650  &   -1.148000  &    2455376.6470   &   0.034407  &   -1.115000  &     2455376.6760   &   0.098370   &  -1.339000   \\
  2455376.6210   &   0.977062  &   -1.136000  &    2455376.6490   &   0.038819  &   -1.128000  &     2455376.6790   &   0.104987   &  -1.371000   \\
  2455376.6230   &   0.981473  &   -1.125000  &    2455376.6520   &   0.045435  &   -1.136000  &     2455376.6820   &   0.111604   &  -1.375000   \\
  2455376.6260   &   0.988090  &   -1.111000  &    2455376.6540   &   0.049847  &   -1.158000  &     2455376.6940   &   0.138071   &  -1.452000   \\
  2455376.6280   &   0.992501  &   -1.099000  &    2455376.6560   &   0.054258  &   -1.178000  &     2455376.6970   &   0.144688   &  -1.441000   \\
  2455376.6310   &   0.999118  &   -1.076000  &    2455376.6590   &   0.060875  &   -1.195000  &     2455376.7010   &   0.153510   &  -1.465000   \\
  2455376.6330   &   0.003529  &   -1.085000  &    2455376.6640   &   0.071903  &   -1.234000  &     2455376.7040   &   0.160127   &  -1.475000   \\
  2455376.6350   &   0.007940  &   -1.090000  &    2455376.6660   &   0.076314  &   -1.260000  &     2455376.7070   &   0.166744   &  -1.480000   \\
  2455376.6380   &   0.014557  &   -1.097000  &    2455376.6690   &   0.082931  &   -1.295000  &     2455376.7100   &   0.173361   &  -1.491000   \\
\hline
\end{tabular}
\end{center}
\textbf
{\footnotesize Notes.} \footnotesize These are only a few lines of the data. The whole data are listed in Table 6, Table 7, Table 8 and Table 9, available as Supporting Information in the online version of this article.
\end{table}

To improve the legibility of the figure, the light curves' data have been shifted vertically which will make no difference to the solutions of W-D program as differential photometry method is used. As shown in Fig. 1, the light curves vary continuously and have very small magnitude differences between the depth of the primary and secondary minima which indicates that V502 Oph is a typical W UMa type contact binary with EW type light curves.

\begin{figure}[!h]
\begin{center}
\includegraphics[width=14cm]{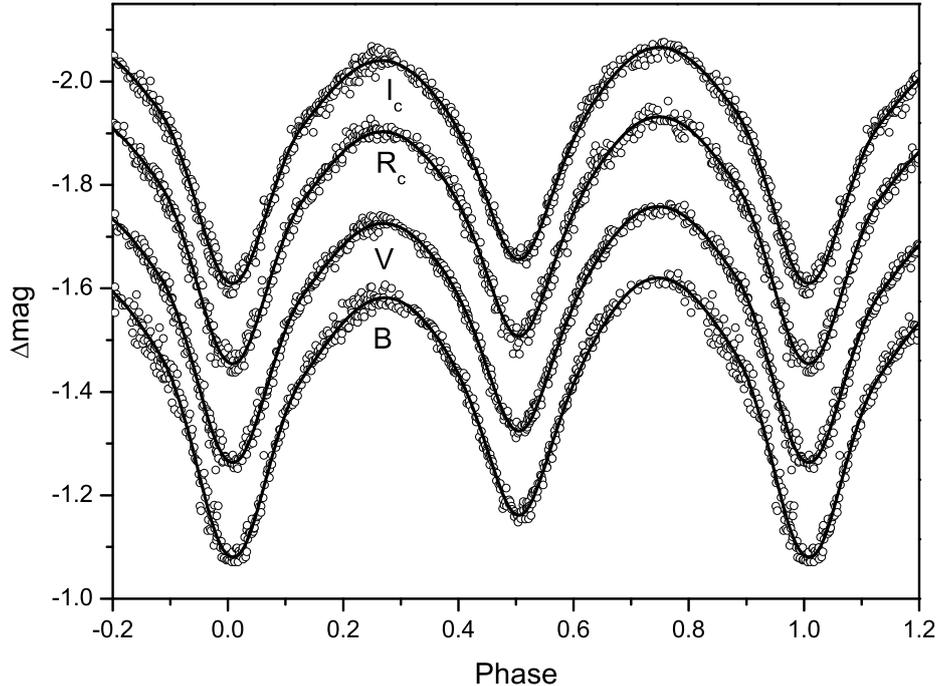}
\caption{The open circles stand for the observational light curves, and the solid lines represent the theoretical light curves determined by the W - D program. The standard deviations for the fitting residuals are 0.016 mag ($B$), 0.011 mag ($V$), 0.015 mag ($R_c$) and 0.013 mag ($I_c$), respectively.}
\end{center}
\end{figure}

\section{Orbital Period Investigation of V502 Oph}

Since it was discovered by \citet{1935AN....255..401H}, V502 Oph has been monitored for nearly eighty years. We have abundant light minima for analyzing the $O - C$ (observed - calculated) curve,  which will reveal dynamic interactions between its components. As times of light minimum determined by visual observations have very large uncertainty, only those obtained by photoelectric observations and CCD camera are collected in our work. Using the following linear ephemeris determined by \citet{1992AJ....103.1658D},
\begin{equation}
Min.I(HJD)=2439639.9562+0^{d}.45339394\times{E},\label{linear ephemeris}
\end{equation}
we obtain the $(O - C)_1$ values and list them in the fourth column of Table \ref{Minimum}. We adopt a quadratic curve to fit the $(O - C)_1$ values, then the residuals ($(O - C)_2$ values) are obtained. The $(O-C)_2$ values of V502 Oph is listed in the fifth column of Table \ref{Minimum} and shown in the middle panel of Fig. 2, which indicates a cyclic change in its orbital period. Thus, a sinusoidal term is superimposed to the ephemeris. The revised ephemeris is
\begin{equation}
\begin{array}{lll}
Min. I=2439639.94876(\pm0.00005)+0.453392912(\pm0.000000005)\times{E}
        \\-1.047(\pm0.002)\times{10^{-10}}\times{E^{2}}\\+0.00402(\pm0.00005)\sin[0.0239(\pm0.0002)\times{E}+263.1(\pm0.7)]
\end{array}
\end{equation}

According to the quadratic term included in the new ephemeris, we conclude that the period of V502 Oph is decreasing continuously at a rate of
$dP/dt=-1.69\times{10^{-7}}day\cdot year^{-1}$. The period of the cyclic change is $P_3=18.7$ years, and its amplitude is $A_3 = 0.00402$ days. The residuals from equation (3) are showed in the bottom panel of Fig. 2. The large scatter of the $(O - C)_2$ values in Fig. 2 may due to the fact that V502 Oph is an active binary system. However, the tendency of the $(O - C)_2$ curve still shows periodic variation obviously and the $(O - C)_2$ curve covers over two circles which makes our results more reliable. We just give the most probable fitting result according to the minimum data available in the bibliography. To determine the period of the third component more accurate, more times of light minimum are need in the future.

\begin{table}[!h]
\caption{$(O-C)$ values of light minima for V502 Oph.}\label{Minimum}
\begin{center}
\small
\begin{tabular}{ccllrrcc}\hline\hline
  JD(Hel.)      &  p/s       &        Epoch        &     $(O-C)_1$   &      $(O-C)_2$   &   Error   &   Method    & Ref.  \\
 (2400000+)     &            &                     &                 &                  &           &             &        \\\hline
34514.7822	    &  p	     &        -11304       &     -0.0089     &      0.0003      &   0.0010  &   pe        & 1      \\
34515.4617	    &  s	     &        -11302.5     &     -0.0095     &     -0.0003      &   0.0008  &   pe        & 2      \\
34840.5465	    &  s	     &        -10585.5     &     -0.0082     &      0.0001      &   0.0003  &   pe        & 2      \\
34897.4501	    &  p	     &        -10460       &     -0.0055     &      0.0026      &           &   pe        & 3      \\
34899.4890	    &  s	     &        -10455.5     &     -0.0069     &      0.0013      &           &   pe        & 3      \\
35257.4459	    &  p	     &        -9666        &     -0.0045     &      0.0028      &   0.0004  &   pe        & 4      \\
35261.5260	    &  p	     &        -9657        &     -0.0049     &      0.0023      &   0.0009  &   pe        & 4      \\
35262.4322	    &  p	     &        -9655        &     -0.0055     &      0.0018      &   0.0005  &   pe        & 4      \\
35544.6726	    &  s	     &        -9032.5      &     -0.0028     &      0.0039      &   0.0003  &   pe        & 4      \\
35579.5852	    &  s	     &        -8955.5      &     -0.0016     &      0.0051      &   0.0004  &   pe        & 4      \\
35582.5274	    &  p	     &        -8949        &     -0.0064     &      0.0002      &   0.0008  &   pe        & 4      \\
35587.5148	    &  p	     &        -8938        &     -0.0064     &      0.0002      &   0.0005  &   pe        & 4      \\
35617.4408	    &  p	     &        -8872        &     -0.0044     &      0.0022      &   0.0012  &   pe        & 4      \\
35621.5205	    &  p	     &        -8863        &     -0.0052     &      0.0013      &   0.0003  &   pe        & 4      \\
37394.9740	    &  s	     &        -4951.5      &     -0.0021     &      0.0028      &           &   pe        & 5      \\
37404.9460	    &  s	     &        -4929.5      &     -0.0048     &      0.0001      &           &   pe        & 5      \\
37405.8546	    &  s	     &        -4927.5      &     -0.0030     &      0.0020      &           &   pe        & 5      \\
39631.7829	    &  p	     &        -18          &     -0.0122     &     -0.0048      &           &   pe        & 6      \\
39637.9078	    &  s	     &        -4.5         &     -0.0081     &     -0.0007      &           &   pe        & 6      \\
39639.9436	    &  p	     &        0            &     -0.0126     &     -0.0052      &           &   pe        & 6      \\
39642.8951	    &  s	     &        6.5          &     -0.0082     &     -0.0007      &           &   pe        & 6      \\
40719.4714	    &  p	     &        2381         &     -0.0158     &     -0.0053      &           &   CCD       & 7      \\
40778.4110	    &  p	     &        2511         &     -0.0174     &     -0.0067      &           &   CCD       & 7      \\
41174.2288	    &  p	     &        3384         &     -0.0125     &     -0.0004      &           &   pe        & 8      \\
41184.2031	    &  p	     &        3406         &     -0.0129     &     -0.0007      &           &   pe        & 8      \\
41194.1799	    &  p	     &        3428         &     -0.0107     &      0.0015      &           &   pe        & 8      \\
41214.1326	    &  p	     &        3472         &     -0.0074     &      0.0049      &           &   pe        & 8      \\
41491.3772	    &  s	     &        4083.5       &     -0.0132     &      0.0002      &           &   CCD       & 9      \\
41802.6320	    &  p	     &        4770         &     -0.0133     &      0.0014      &           &   pe        & 10     \\
41833.4620	    &  p	     &        4838         &     -0.0141     &      0.0008      &           &   CCD       & 11     \\
41844.3440	    &  p	     &        4862         &     -0.0135     &      0.0014      &           &   CCD       & 11     \\
41858.4000	    &  p	     &        4893         &     -0.0128     &      0.0022      &           &   CCD       & 11     \\
41860.4400	    &  s	     &        4897.5       &     -0.0130     &      0.0020      &           &   CCD       & 11     \\
41888.3200	    &  p	     &        4959         &     -0.0168     &     -0.0016      &           &   CCD       & 9      \\
41888.3205	    &  p	     &        4959         &     -0.0163     &     -0.0011      &           &   pe        & 9      \\
42570.4558	    &  s	     &        6463.5       &     -0.0121     &      0.0063      &           &   CCD       & 12     \\
43238.5240	    &  p	     &        7937         &     -0.0199     &      0.0023      &           &   CCD       & 13     \\
43238.5250	    &  p	     &        7937         &     -0.0189     &      0.0033      &           &   CCD       & 13     \\
43292.4761	    &  p	     &        8056         &     -0.0217     &      0.0008      &           &   CCD       & 13     \\
43665.8463	    &  s	     &        8879.5       &     -0.0214     &      0.0034      &           &   pe        & 14     \\
\hline
\end{tabular}
\end{center}
\end{table}

\addtocounter{table}{-1}
\begin{table}[!h]
\small
\begin{center}
\caption{Continuted}
\begin{tabular}{ccllrrcc}\hline\hline
  JD(Hel.)      &  p/s       &        Epoch        &     $(O-C)_1$   &      $(O-C)_2$   &   Error   &   Method    & Ref.  \\
 (2400000+)     &            &                     &                 &                  &           &             &        \\\hline
43666.7545	    &  s	     &        8881.5       &     -0.0200     &      0.0049      &           &   pe        & 14     \\
43668.7951	    &  p	     &        8886         &     -0.0197     &      0.0052      &           &   pe        & 14     \\
43671.7469	    &  s	     &        8892.5       &     -0.0149     &      0.0100      &           &   pe        & 14     \\
44370.8684	    &  s	     &        10434.5      &     -0.0269     &      0.0027      &   0.0007  &   CCD       & 15     \\
46528.5488	    &  s	     &        15193.5      &     -0.0482     &     -0.0010      &   0.0005  &   pe        & 16     \\
46529.4572	    &  s	     &        15195.5      &     -0.0466     &      0.0006      &   0.0005  &   pe        & 16     \\
46530.5866	    &  p	     &        15198        &     -0.0507     &     -0.0034      &   0.0004  &   pe        & 16     \\
46552.8030	    &  p	     &        15247        &     -0.0506     &     -0.0031      &           &   pe        & 17     \\
46555.5244	    &  p	     &        15253        &     -0.0496     &     -0.0021      &   0.0004  &   pe        & 16     \\
46582.4954	    &  s	     &        15312.5      &     -0.0555     &     -0.0078      &           &   pe        & 18     \\
46584.5391	    &  p	     &        15317        &     -0.0521     &     -0.0043      &           &   pe        & 18     \\
46585.4478	    &  p	     &        15319        &     -0.0502     &     -0.0024      &           &   pe        & 18     \\
46586.3496	    &  p	     &        15321        &     -0.0552     &     -0.0074      &           &   pe        & 18     \\
46586.3496	    &  p	     &        15321        &     -0.0552     &     -0.0074      &           &   pe        & 18     \\
46586.5744	    &  s	     &        15321.5      &     -0.0571     &     -0.0093      &           &   pe        & 18     \\
46587.4811	    &  s	     &        15323.5      &     -0.0571     &     -0.0094      &           &   pe        & 18     \\
46590.4235	    &  p	     &        15330        &     -0.0618     &     -0.0140      &           &   pe        & 18     \\
46595.4205	    &  p	     &        15341        &     -0.0521     &     -0.0043      &           &   pe        & 18     \\
46640.5356	    &  s	     &        15440.5      &     -0.0497     &     -0.0015      &           &   pe        & 17     \\
46915.5165	    &  p	     &        16047        &     -0.0523     &     -0.0014      &   0.0004  &   pe        & 16     \\
46961.7750	    &  p	     &        16149        &     -0.0399     &      0.0114      &           &   pe        & 19     \\
47707.3612	    &  s	     &        17793.5      &     -0.0601     &     -0.0012      &           &   pe        & 20     \\
48094.3284	    &  p	     &        18647        &     -0.0646     &     -0.0016      &           &   pe        & 20     \\
48096.3673	    &  s	     &        18651.5      &     -0.0660     &     -0.0029      &           &   pe        & 20     \\
48099.3150	    &  p	     &        18658        &     -0.0653     &     -0.0023      &           &   pe        & 20     \\
48114.2770	    &  p	     &        18691        &     -0.0653     &     -0.0021      &           &   pe        & 20     \\
51266.9174	    &  s	     &        25644.5      &     -0.0997     &      0.0030      &   0.0002  &   CCD       & 21     \\
51267.1421	    &  p	     &        25645        &     -0.1017     &      0.0010      &   0.0003  &   CCD       & 21     \\
51299.1037	    &  s	     &        25715.5      &     -0.1044     &     -0.0012      &           &   CCD       & 22     \\
51299.1040	    &  s	     &        25715.5      &     -0.1041     &     -0.0009      &   0.0011  &   CCD       & 23     \\
51319.0560	    &  s	     &        25759.5      &     -0.1014     &      0.0020      &   0.0010  &   CCD       & 23     \\
51320.4155	    &  s	     &        25762.5      &     -0.1021     &      0.0013      &   0.0002  &   CCD       & 21     \\
51320.6436	    &  p	     &        25763        &     -0.1007     &      0.0028      &   0.0004  &   CCD       & 21     \\
51388.8781	    &  s	     &        25913.5      &     -0.1020     &      0.0024      &   0.0005  &   CCD       & 21     \\
51389.1042	    &  p	     &        25914        &     -0.1026     &      0.0018      &   0.0002  &   CCD       & 21     \\
51666.1193	    &  p	     &        26525        &     -0.1112     &     -0.0028      &           &   CCD       & 24     \\
52043.1141	    &  s	     &        27356.5      &     -0.1134     &      0.0005      &           &   CCD       & 25     \\
52768.5242	    &  s	     &        28956.5      &     -0.1336     &     -0.0086      &   0.0002  &   CCD       & 26     \\
52772.3815	    &  p	     &        28965        &     -0.1302     &     -0.0051      &   0.0002  &   CCD       & 27     \\
53189.0452	    &  p	     &        29884        &     -0.1355     &     -0.0038      &           &   CCD       & 28     \\
\hline
\end{tabular}
\end{center}
\end{table}

\addtocounter{table}{-1}
\begin{table}[!h]
\small
\begin{center}
\caption{Continuted}
\begin{tabular}{ccllrrcc}\hline\hline
  JD(Hel.)      &  p/s       &        Epoch        &     $(O-C)_1$   &      $(O-C)_2$   &   Error   &   Method    & Ref.  \\
 (2400000+)     &            &                     &                 &                  &           &             &        \\\hline
53402.5929	    &  p	     &        30355        &     -0.1364     &     -0.0012      &   0.0003  &   CCD       & 29     \\
53405.5412	    &  s	     &        30361.5      &     -0.1351     &      0.0001      &   0.0008  &   CCD       & 29     \\
53500.5220	    &  p	     &        30571        &     -0.1403     &     -0.0036      &   0.0009  &   CCD       & 30     \\
53524.3252	    &  s	     &        30623.5      &     -0.1403     &     -0.0032      &   0.0004  &   CCD       & 30     \\
53537.4733	    &  s	     &        30652.5      &     -0.1407     &     -0.0033      &   0.0003  &   CCD       & 31     \\
53863.4527	    &  s	     &        31371.5      &     -0.1515     &     -0.0087      &   0.0002  &   CCD       & 32     \\
53905.3952	    &  p	     &        31464        &     -0.1479     &     -0.0045      &   0.0003  &   CCD       & 31     \\
54263.1223	    &  p	     &        32253        &     -0.1487     &      0.0009      &           &   CCD       & 33     \\
54577.5419	    &  s	     &        32946.5      &     -0.1577     &     -0.0028      &   0.0006  &   CCD       & 34     \\
54613.1335	    &  p	     &        33025        &     -0.1576     &     -0.0020      &           &   CCD       & 35     \\
54631.0408	    &  s	     &        33064.5      &     -0.1593     &     -0.0034      &           &   CCD       & 35     \\
54641.4712	    &  s	     &        33087.5      &     -0.1570     &     -0.0009      &   0.0006  &   CCD       & 34     \\
55661.5923	    &  s	     &        35337.5      &     -0.1723     &      0.0023      &   0.0003  &   CCD       & 36     \\
55700.1253	    &  s	     &        35422.5      &     -0.1777     &     -0.0025      &           &   CCD       & 37     \\
56092.9876	    &  p	     &        36289        &     -0.1813     &      0.0014      &           &   CCD       & 38     \\
56428.4975	    &  p	     &        37029        &     -0.1829     &      0.0062      &   0.0003  &   CCD       & 36     \\
56447.0859	    &  p	     &        37070        &     -0.1837     &      0.0058      &           &   CCD       & 39     \\
56447.0860	    &  p	     &        37070        &     -0.1836     &      0.0059      &           &   CCD       & 39     \\
56447.0862	    &  p	     &        37070        &     -0.1834     &      0.0061      &           &   CCD       & 39     \\
56452.0720	    &  p	     &        37081        &     -0.1849     &      0.0047      &           &   CCD       & 39     \\
56460.4572	    &  s	     &        37099.5      &     -0.1875     &      0.0022      &   0.0006  &   CCD       & 36     \\\hline
\end{tabular}
\end{center}
\textbf
{\footnotesize Reference:} \footnotesize (1) \citet{1964AJ.....69..316F}; (2) \citet{1958BAN....14..131K}; (3) \citet{1962AcA....12..181S}; (4) \citet{1968BANS....2..277K}; (5) \citet{1967AJ.....72.1028W}; (6) \citet{1969AJ.....74..218B}; (7) \citet{1971IBVS..508....1P}; (8) \citet{1975PZP.....2..161P}; (9) \citet{1974IBVS..937....1K}; (10) \citet{1976PASP...88..531S}; (11) \citet{1973IBVS..842....1V}; (12) \citet{1976IBVS.1163....1P}; (13) \citet{1978IBVS.1449....1E}; (14) \citet{1979IBVS.1569....1M}; (15) \citet{1984IBVS.2545....1S}; (16) \citet{1988IBVS.3218....1Z}; (17) \citet{1987AJ.....94..792L}; (18) \citet{1988AAS...74..265R}; (19) \citet{1989RMxAA..17...97H}; (20) \citet{1992AJ....103.1658D}; (21) \citet{2004IBVS.5507....1O}; (22) \citet{2000VSB.No.37}; (23) \citet{2008VSB.No.47}; (24) \citet{2001VSB.No.38}; (25) \citet{2002VSB.No.39};
(26) \citet{2005IBVS.5662....1B}; (27) \citet{2005IBVS.5616....1B}; (28) \citet{2005VSB.No.43}; (29) \citet{2007IBVS.5809....1S}; (30) \citet{2005IBVS.5649....1A}; (31) \citet{2007IBVS.5754....1S}; (32) \citet{2006IBVS.5707....1D}; (33) \citet{2008VSB.No.46}; (34) \citet{2009IBVS.5887....1Y}; (35) \citet{2009VSB.No.48}; (36) \citet{2013OEJV..160....1H};
(37)\citet{2012VSB.No.53}; (38) \citet{2013VSB.No.55}; (39) \citet{2014VSB.No.56};
\end{table}

\begin{figure}[!h]
\begin{center}
\includegraphics[width=13cm]{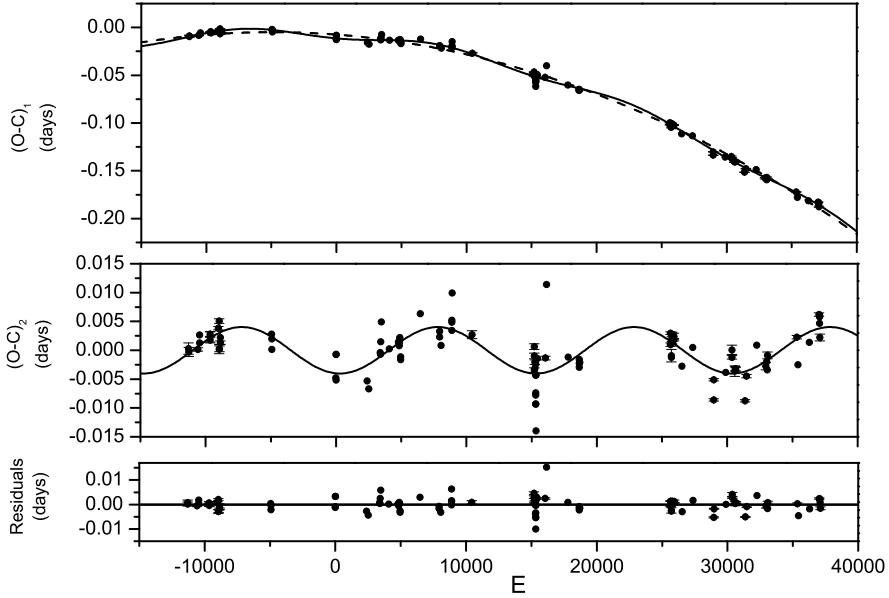}
\caption{The $(O-C)_1$ values of V502 Oph calculated from Equation (2) is dispalyed in the upper panel. The solid line represents the combination of a parabolic variation and a periodic change, and the dash line refers  to the downward parabolic variation. In the middle of Fig. 2, the $(O-C)_2$ values are shown while the quadratic part in Equation(3) are subtracted, where a cyclic change is more obviously to be seen. The fitting residuals from Equation(3) were plotted in the bottom panel.}
\end{center}
\end{figure}

\section{Photometric Solutions of V502 Oph}

In the present paper, a simultaneous multi-color light curves' analysis of V502 Oph is performed with the Wilson-Devinney (W-D) program (Version 2013) \citep{Wilson1971,Van2007,Wilson2010} to derive its physical parameters. According to the Radial Velocity (RV) studies published by \citet{2004AJ....127.1712P}, the mass ratio (q) of V502 Oph is fixed at $q = 2.985$.
The effective temperature of the hotter component ($T1$) in V502 Oph is fixed to $T1 = 6140K$, which was used by \citet{1988AAS...74..265R} and \citet{1988IBVS.3218....1Z} corresponding to its spectral type of F9V \citep{1959ApJ...130..789S}. V502 Oph is a late-type W-subtype W UMa contact binary system, so convective envelopes is assumed. The gravity-darkening coefficients and bolometric albedo adoptable for convective envelopes are used, which were $g_1=g_2=0.32$ \citep{1967ZA.....65...89L} and $A_1=A_2=0.5$ \citep{1969AcA....19..245R}. And we choose the limb darkening coefficients from \citet{1993AJ....106.2096V}'s table in logarithmic law.

While modeling the light curves of V502 Oph, mode 3 for contact system is chosen. The adjustable parameters are: the orbital inclination of V502 Oph ($i$); the surface temperature for the secondary star ($T_{2}$); the monochromatic luminosity of the primary star ($L_{1B}$, $L_{1V}$, $L_{1R}$ and $L_{1I}$); the dimensionless surface potential of the primary and secondary star ($\Omega_{1}=\Omega_{2}$ for contact binary ). Since both spectroscopic observations and orbital period variations confirmed the existence of a tertiary component around V502 Oph, the third light ($l_3$) is also set as an adjustable parameter in our solutions. It has to be mentioned that we use a spot model in the calculating as the O'Connell effect appears in the light curves. The final parameters of the solutions are list in Table \ref{phsolutions}. The theoretical light curves are displayed in Fig. 1.

\begin{table}[!h]
\begin{center}
\caption{Photometric solutions of V502 Oph}\label{phsolutions}
\small
\begin{tabular}{lllllllll}
\hline\hline
Parameters                            &    Values                    &    Values    \\
                                      &  with $l_3$                  &  without $l_3$     \\\hline
$g_{1}$                               &  0.32(fixed)                 &  0.32(fixed)   \\
$g_{2}$                               &  0.32(fixed)                 &  0.32(fixed)   \\
$A_{1}$                               &  0.50(fixed)                 &  0.50(fixed)    \\
$A_{2}$                               &  0.50(fixed)                 &  0.50(fixed)    \\
q ($M_2/M_1$ )                        &  2.985(fixed)                &  2.985(fixed)    \\
$T_{1}(K)   $                         &  6140(fixed)                 &  6140(fixed)     \\
$i(^{\circ})$                         &  76.4($\pm0.6$)              &  72.8($\pm0.1$)   \\
$\Omega_{in}$                         &  6.6030                      &  6.6030  \\
$\Omega_{out}$                        &  5.9843                      &  5.9843  \\
$\Omega_{1}=\Omega_{2}$               &  6.3779($\pm0.0125$)         &  6.4384($\pm0.0056$)\\
$T_{2}(K)$                            &  5900($\pm8$)                &  5922($\pm6$)\\
$\Delta T(K)$                         &  240                         &  218 \\
$T_{2}/T_{1}$                         &  0.961($\pm0.001$)           &  0.964($\pm0.001$)\\
$L_{1}/(L_{1}+L_{2}$) ($B$)           &  0.3285($\pm0.0033$)         &  0.3206($\pm0.0010$) \\
$L_{1}/(L_{1}+L_{2}$) ($V$)           &  0.3149($\pm0.0050$)         &  0.3085($\pm0.0009$)  \\
$L_{1}/(L_{1}+L_{2}$) ($R_c$)         &  0.3088($\pm0.0049$)         &  0.3030($\pm0.0008$) \\
$L_{1}/(L_{1}+L_{2}$) ($I_c$)         &  0.3038($\pm0.0047$)         &  0.2985($\pm0.0007$)  \\
$L_{1}/(L_{1}+L_{2}+L_{3}$) ($B$)     &  0.2886($\pm0.0062$)         &        \\
$L_{1}/(L_{1}+L_{2}+L_{3}$) ($V$)     &  0.2742($\pm0.0076$)         &         \\
$L_{1}/(L_{1}+L_{2}+L_{3}$) ($R_c$)   &  0.2697($\pm0.0074$)         &          \\
$L_{1}/(L_{1}+L_{2}+L_{3}$) ($I_c$)   &  0.2632($\pm0.0071$)         &         \\
$L_{3}/(L_{1}+L_{2}+L_{3}$) ($B$)     &  0.1215($\pm0.0163$)         &         \\
$L_{3}/(L_{1}+L_{2}+L_{3}$) ($V$)     &  0.1294($\pm0.0192$)         &          \\
$L_{3}/(L_{1}+L_{2}+L_{3}$) ($R_c$)   &  0.1264($\pm0.0189$)         &        \\
$L_{3}/(L_{1}+L_{2}+L_{3}$) ($I_c$)   &  0.1335($\pm0.0183$)         &        \\
$r_{1}(pole)$                         &  0.2854($\pm0.0010$)         &  0.2805($\pm0.0004$)  \\
$r_{1}(side)$                         &  0.2998($\pm0.0012$)         &  0.2939($\pm0.0005$)   \\
$r_{1}(back)$                         &  0.3469($\pm0.0022$)         &  0.3360($\pm0.0009$)  \\
$r_{2}(pole)$                         &  0.4619($\pm0.0009$)         &  0.4577($\pm0.0004$)  \\
$r_{2}(side)$                         &  0.4992($\pm0.0012$)         &  0.4935($\pm0.0005$)  \\
$r_{2}(back)$                         &  0.5306($\pm0.0016$)         &  0.5231($\pm0.0007$) \\
$f$                                   &  $35.3\,\%$($\pm$2.0\,\%$$)  &  $25.5\,\%$($\pm$0.9\,\%$$) \\
$\theta(^{\circ})$                    &  102.6($\pm14.4$)            &  106.0($\pm10.9$)      \\
$\psi(^{\circ})$                      &  116.4($\pm1.5$)             &  116.6($\pm1.5$)  \\
$r$(rad)                              &  0.36($\pm0.12$)             &  0.36($\pm0.12$)\\
$T_f$                                 &  0.87($\pm0.09$)             &  0.87($\pm0.08$)  \\
$\Sigma{\omega(O-C)^2}$               &  0.055668                    &  0.066069\\
\hline
\hline
\end{tabular}
\end{center}
\textbf
{\footnotesize Notes.} \footnotesize $L_3 = 4 \times \pi \times l_3$ for condisering a isotropically radiation. The spot parameters pertain to a spot located on the larger, more massive star.
\end{table}

\section{Discussions and Conclusions}

We have present multi-color ($B$ $V$ $R_c$ $I_c$) CCD photometric solutions of the contact binary V502 Oph. The light curves have a long-lived asymmetric phenomenon \citep{2015IBVS.6127....1K} in the placement of the light maxima called O'Connell effect \citep{1969CoKon..65..457M}. We use a spot model to fit the light curves in the W-D program. The solutions indicate that V502 Oph is a W-subtype contact binary with the contact degree of $f = 35.3\,\%(\pm2.0\,\%)$. The temperature difference between its two components is $240K$ and the more massive star has a lower surface temperature. Combining the orbital inclination ($i = 76.4^{\circ}\pm0.6$) of ours and the mass function given by \citet{2004AJ....127.1712P}: $(M_1+M_2)sin^3i = 1.679\pm0.022M_\odot$, the absolute physical parameters of the two components in V502 Oph are determined and listed in Table \ref{absolute}. The orbital semi-major axis of V502 Oph is calculated to be $a = 3.04(\pm0.02)R_\odot$. We have also checked the degeneracy of the third light, and the results are also in Table \ref{absolute}. It is found that degeneracy of the third light result in very little difference on parameters such as $i$, $T_{2}$, $\Omega_{1}$. However, the solutions with the third light give out a much smaller fitting residual.

\begin{table}[!h]
\caption{Absolute parameters of the two components in V502 Oph}\label{absolute}
\begin{center}
\small
\begin{tabular}{lllllllll}
\hline
Parameters                        &Primary                         & Secondary          \\
\hline
$M$                               & $0.46(\pm0.02)M_\odot$         & $1.37(\pm0.02)M_\odot$         \\
$R$                               & $0.94(\pm0.01)R_\odot$         & $1.51(\pm0.01)R_\odot$         \\
$L$                               & $1.13(\pm0.02)L_\odot$         & $2.49(\pm0.03)L_\odot$         \\
\hline
\end{tabular}
\end{center}
\end{table}

The light curves of V502 Oph have been investigated by many other researchers (Table 5). However, there is not any information about third light $(l_3)$ in their solutions. As shown in Table 5, \citet{2011MNRAS.412.1787D} assumed that the hotter component is a G0V type star and set the temperature to be $T1 = 5940K$. Only \citet{1998vsr..conf..184R} used spot model in their solutions. In our solutions, we set the temperature to be $T1 = 6140K$ according to its spectral type of F9V and add a spot in our model considering the light curves' asymmetry. Compared with the solutions listed in Table 5, we obtain a higher orbital inclination and contact degree, which may be caused by the contribution of third light in our solutions. 

\begin{table}[!h]
\begin{center}
\caption{Photometric solutions of V502 Oph published by other researchers}
\tiny
\begin{tabular}{cccccccccccc}
\hline\hline
Parameters                  & \citet{1982AAS...49..123M}  & \citet{1988IBVS.3218....1Z} & \citet{1988AAS...74..265R}  & \citet{1998vsr..conf..184R}& \citet{2011MNRAS.412.1787D}  \\\hline
$T_{1}(K)   $               & 6200                        & 6140                        & 6140                        & 6200                       & 5940                       \\
q ($M_2/M_1$ )              & 2.65                        & 2.64                        &                             & 2.70                       & 2.985          \\
$i(^{\circ})$               & 71.3($\pm0.5$)              & 70.2($\pm0.2$)              & 68.5($\pm0.5$)              & 71.0($\pm0.6$)             & 69.77($\pm0.21$)        \\
$\Omega_{1}=\Omega_{2}$     & 6.004($\pm0.026$)           & 6.072($\pm0.025$)           &                             &                            & 6.397($\pm0.013$)       \\
$f$                         & $23.6\,\%$                  & $10.3\,\%$($\pm$4.1\,\%$$)  &                             &                            & $22\,\%$         \\
$T_{2}(K)$                  & 5968($\pm35$)               & 5750($\pm30$)               & 5800                        & 5960                       & 5739($\pm129$)              \\
$\Delta T(K)$               & 232                         & 390                         & 340                         & 240                        & 201                         \\
$T_{2}/T_{1}$               & 0.9626($\pm0.0056$)         & 0.9365($\pm0.0049$)         & 0.9446                      & 0.9613                     & 0.966($\pm0.022$)        \\
$L_{1}/(L_{1}+L_{2})$ ($B$) & 0.338($\pm0.006$)           & 0.3732($\pm0.0004$)         & 0.370($\pm0.010$)           & 0.269($\pm0.010$)          &                          \\
$L_{1}/(L_{1}+L_{2})$ ($V$) & 0.334($\pm0.005$)           & 0.3581($\pm0.0004$)         & 0.361$\pm0.010$)            &                            & 0.3063($\pm0.0329$)                     \\
$\theta(^{\circ})$          &                             &                             &                             & 90 (star 1), 30 (star 2)   &                         \\
$\psi(^{\circ})$            &                             &                             &                             & 180 (star 1), 270 (star 2) &                      \\
$r$(rad)                    &                             &                             &                             & 20 (star 1), 40 (star 2)   &                          \\
$T_f$                       &                             &                             &                             & 0.7 (star 1), 0.7 (star 2) &                         \\
\hline
\end{tabular}
\end{center}
\end{table}

W UMa type contact binaries have been observed and analyzed for over half century. However, there is not a generally accepted formation and evolution scenario for them. Numerical calculations suggest that W UMa type contact binaries are formed from initially detached binary systems. The initially detached systems are mainly controlled by two mechanisms in the pre-contact phase, namely nuclear evolution of the initially massive star and angular momentum loss of the initially lighter star \citep{2006AcA....56..199S,2011AcA....61..139S,2014MNRAS.437..185Y}. Considering the physical parameters listed in Table \ref{absolute} and the method established by \citet{2013MNRAS.430.2029Y}, the initial mass of the two components in V502 Oph are calculated to be $M_{1i} = 1.87(\pm0.01)M_\odot$ and $M_{2i} = 0.86(\pm0.04)M_\odot$. The initial mass of the present primary one ($M_{2i}$) indicates that it is a late-type star, which has played a very important role in early angular momentum evolution of close binaries as magnetic braking is supposed to be the main mechanism for such stars with convective envelope. The mass loss of the initially massive star is calculated to be $1.41M_\odot$ and the mass gained by the present massive star is $0.49M_\odot$. About one-third of mass lost by the initially massive star is transferred to the present massive star.

The $O-C$ analysis shows that the orbital period of V502 Oph is decreasing at a rate of $dP/dt=-1.69\times{10^{-7}}day\cdot year^{-1}$. It may be caused by the mass transfer from the more massive component to the less massive one. By assuming a conservative mass transfer case and using the well-known equation
\begin{equation}
\begin{array}{lll}
 \frac{dM_{2}}{dt}=- \frac{M_1M_2}{3P(M_1-M_2)}\times{dP/dt},
 \end{array}
\end{equation}
the mass transfer is determined to be $\frac{dM_{2}}{dt}= - 3.01\times{10^{-8}}M_\odot/year$. However, the real mass transfer rate may be quite different while the contribution of angular momentum loss (AML) is considered.

As shown in Fig. 2, after the downward parabolic variation in the $(O-C)_{1}$ curve is subtracted, the $(O-C)_{2}$ residuals suggest that there is a cyclic variation with a period of 18.7 years. It is due to the the light-travel time effect that caused by the gravitational influence of a tertiary component. By assuming a circular orbit, the projected radius ($a'_{12}\sin i'$) of the orbit that the eclipsing binary rotates around the barycenter of the triple system is calculated with the following equation,
\begin{equation}
\begin{array}{lll}
a'_{12}\sin i'=A_3 \times c,
 \end{array}
\end{equation}
where $A_3$ is the amplitude of the periodic change and $c$ is the speed of light. The values is calculated to be $a'_{12}\sin i'=0.69(\pm0.01)AU$. The mass function and the mass of the tertiary companion are computed with the following equation,
 \begin{equation}
\begin{array}{lll}
f(m)=\frac{4\pi^2}{GP^2_3}\times(a'_{12}\sin i')^3=\frac{(M_3\sin i')^3}{(M_1+M_2+M_3)^2},
 \end{array}
\end{equation}
where $G$ is the gravitational constant and $P_3$ is the period that the tertiary component orbits around V502 Oph. The mass function are calculated to be $f(m) = 0.00096M_\odot$. The $i'$-$M_3$ and $i'$-$a_3$ diagram are displayed in Fig. 3. The red star in Fig. 3 are the values of the tertiary component when the orbits of the three components are coplanar.

\begin{figure}[!h]
\begin{center}
\includegraphics[width=14cm]{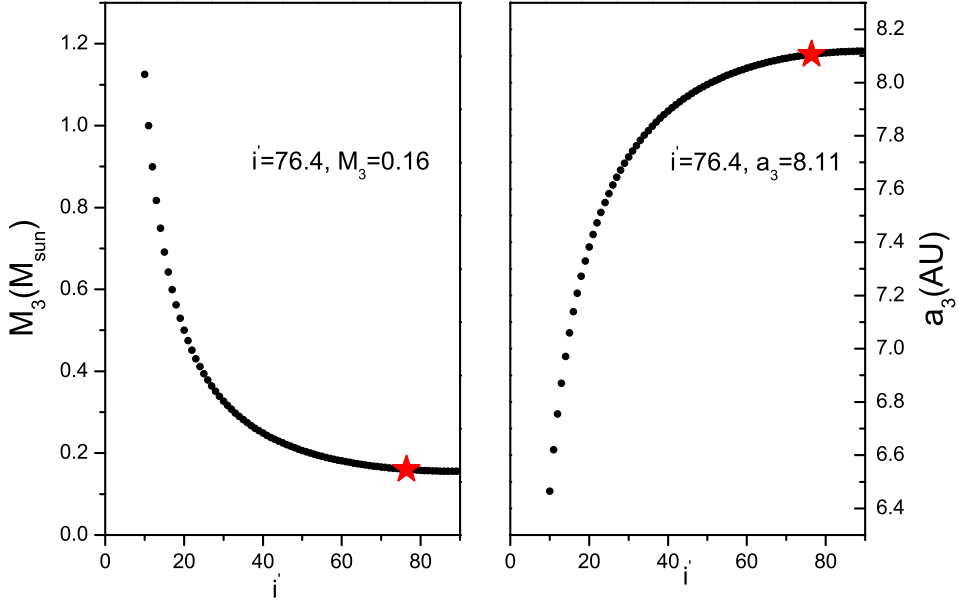}
\caption{Relations between the mass $M_3$ (in units of $M_\odot$) of the third component and its orbital inclination $i'(^{\circ})$, and relations between the orbital radius $a_3$ (in units of AU) and its orbital inclination $i'(^{\circ})$ of the tertiary component  in the V502 Oph system. The red star is the position of the tertiary component when it is co-planer with the binary system.}
\end{center}
\end{figure}

During the photometric solutions, the third light was also included as an adjustable parameters, and our photometric solutions confirm the existence of the third light ($l_3$) in the binary system. Based on the amounts of the third light listed in Table 3, the $B - V$ color of the light source is calculated to be 0.61, which may correspond to be a G1 type main sequence star with its mass to be nearly $1.0M_\odot$. As shown in Fig. 4, the semi-major axes of the tertiary component is less than 8.11 AU and its mass is larger than $0.16 M_\odot$. The tertiary component confirmed by the $O - C$ curve analysis is a stellar type component and it can not be outside the measured area on the CCD image considering the distance to V502 Oph of 86($\pm8$) pc determined by Hipparchos \citep{1997ESASP1200.....E}. Therefore, the tertiary detected by the $O - C$ curve analysis may possibly be the third light source. In this case, the orbital inclination of the tertiary component will less than 20 degrees in Fig. 3.

Although tertiary components are quite common in W UMa systems, the formation theory of triple systems and the effects of the tertiary component on the close binary systems are not clear. The tertiary component around V502 Oph may accelerate the orbital evolution of the binary system by removing angular momentum from it \citep{2013ApJS..209...13Q}. As for V502 Oph, its close-in companion has been confirmed and comprehensively researched through light curves' solutions, $O - C$ curve analysis and spectroscopic observations, which make it an important target for testing formation theories of W UMa-type binaries, and dynamical interaction and angular momentum evolution of multiple systems.

\bigskip

\vskip 0.3in \noindent
This work is supported by the Chinese Natural Science Foundation (Grant No. 11133007, 11325315, 11203066 and No. 11403095), the Strategic Priority Research Program ``The Emergence of Cosmological Structure'' of the Chinese Academy of Sciences (Grant No. XDB09010202) and the Science Foundation of Yunnan Province (Grant No. 2012HC011 and 2014FB187). We thank Koju, Vijay and Beaky, Matthew M. for the high accuracy light curves' data they observed using the 20-cm Meade LX200GPS telescope at the Truman State University Observatory in Kirksville, Missouri.


\end{document}